# Social Synergetics, Social Physics and Research of Fundamental Laws in Social Complex Systems

Yi-Fang Chang

Department of Physics, Yunnan University, Kunming, 650091, China

(e-mail: yifangchang1030@hotmail.com)

**Abstract**: We proposed social synergetics and the four basic theorems, in which theorem of perfect correlation on humanity is researched mathematically. Generally, we discuss the four variables and the eight aspects in social physics. We search social thermodynamics and the five fundamental laws of social complex systems. Then we research different relations among social elements and applications of the nonlinear sociology, for example, for the economic systems. Finally, we discuss the evolutional equation of system and the educational equation.

**1.Social synergetics**

H.Haken proposed synergetics [1-3], which as a bridge between the natural and social sciences and a strategy to cope with various complex systems are very useful. Social synergetics is an application of synergetics [4]. Both basic mathematics and equations are the same [4]. The mathematical base in synergetics is very universal probability, information and some equations, for example [1]:

(1). The master equation with random movement model is:

$$\frac{dP(m,t)}{dt} = W(m,m-1)P(m-1,t) + W(m,m+1)P(m+1,t) - [W(m+1,m) + W(m-1,m)]P(m,t). \tag{1}$$

For the Markov process, Eq.(1) derives the Chapman-Kolmogorov equation

$$P_{t3,t1}(m_3,m_1) = \sum_{m2} P_{t3,t2}(m_3,m_2) P_{t3,t1}(m_2,m_1). \tag{2}$$

Its linearized algebraic equation is:

$$\sum_{n=1}^{N} W(m,n)P(n) - P(m)\sum_{n=1}^{N} W(n,m) = 0. \tag{3}$$

(2).The Langevin equation described the Brownian movement:

$$\frac{dV}{dt} = -\alpha V + F(t). \tag{4}$$

Its solution is:

$$V(t) = \int_0^t exp[-\alpha(t-\tau)]F(\tau)d\tau + V(0)\exp(-\alpha t). \tag{5}$$

(3).The Fokker-Planck equation is:

$$\frac{df}{dt} = \frac{d}{dq}(\gamma q f) + \frac{1}{2}Q\frac{d^2 f}{dq^2}. \tag{6}$$

For the multivariate and any force $K_i(\vec{q})$, if the form of the Langevin equation is:

$$\frac{dq_i}{dt} = K_i(\vec{q}) + F_i(t), \tag{7}$$

corresponding Fokker-Planck equation will be:

$$\frac{df}{dt} = -\nabla_q \{\vec{K}f\} + \frac{1}{2}\sum_{ij} Q_{ij} \frac{\partial^2}{\partial q_i \partial q_j} f. \tag{8}$$

(4).The general Ginzburg-Landau equation of the nonequilibrium phase transition is:

$$\frac{dq}{dt} = -\alpha q - \beta q^3 + \gamma \Delta q + F(t). \tag{9}$$

Universally applicable methods originating in statistical physics and synergetics are combined with concepts from social science in order to set up and to apply a model construction concept for the quantitative description of a broad class of collective dynamical phenomena within society. Weidlich discussed physics and social science as the approach of synergetics [5], and constructed probabilistic transition rates between attitudes and actions from the decisions of individuals and introducing the concept of dynamical utilities. The latter enter the central equation of motion, i.e. the master equation, for the probability distribution over the possible macroconfigurations of society. From the master equation the equations of motion for the expectation values of the macrovariables of society can be derived. These equations are in general nonlinear. Their solutions may include stationary solutions, limit cycles and strange attractors, and with varying trend parameters also phase transitions between different modes of social behaviour. The general model construction approach is subsequently applied to characteristic examples from different social sciences, such as sociology, demography, regional science and economics. These examples refer to collective political opinion formation, to interregional migration of interactive populations, to settlement formation on the micro-, meso- and macroscale, and to nonlinear nonequilibrium economics, including market instabilities.

Synergetics influences relation between science and society, and accords with the cooperation thought in the East-West civilizations. It provides successfully a theoretical pattern transforming from random to order, and derives many brilliant examples in the life science and the social science [1-4]. Among elements the competitiveness derives the slave principle in synergetics, the cooperativeness forms a self-organization.

Some social problems and the global problems may be solved by synergetics as a theoretical base. A fundamental characteristic of social synergetics is a synergy with difference, even opposition. It is a harmony, or cooperation, or consistency passing reasonably consultation and balance in order to exist and develop for total social system, or to benefit majority. The social synergetics researches a general cooperation on different aspects and different levels. It discusses synergy from a world angle and from common benefit of humanity.

In social synergetics [4], we describe mathematically its basic principles, in which the synergy is a necessary condition to the existence of any social system. Determinacy and probability are complementary and unified in production and evolution. The seven rules on development of enterprise are discussed concretely.

In social synergetics we proposed the four basic theorems of social systems [4]:

1.Theorem of outside selection and inside evolution. It includes two types of synergy: outside inclusion and internal cooperation. Here the outside selection is analogue with the social context principle expounded by Xie Yu [6].

2.Theorem of golden section on social synergy. It is a quantitative target of stable existence and synergy on social system.

3.Theorem of perfect correlation on humanity. It points out that everyone in social system correlates each other by the formula. We introduce a fight formula [4] like a Bell inequation [7]:

$$\frac{1}{k}[(ab)+|a-b|] = J < 1, \quad (10)$$

in which 1 as an initial state of the whole, and $a$ and $b$ are its parts. Because $k$ is a fight-strength in this case, J is always less than 0. By mathematical extremal theorem, when $k=1$ and $a=b=1/2$, there is the minimum J=1/4.

Further development can be both forms:

(1).For $a-b$ is not the absolute value, so

$$\frac{1}{k}[(ab)+(a-b)] = J_1. \quad (11)$$

It shows that if fewness fights majority, a result will become to an opposite direction, and $J_1$ may be negative. When $a = (3-\sqrt{5})/2 = 0.382$, $b = (\sqrt{5}-1)/2 = 0.618$ (a golden section value), $J_1 = 0$. While $a<0.382$, $J_1 <0$. This may explain that any despotism society of rule as fewness will tend necessarily to perdition. The Solon reformation in old Green and the imperial examinations in old China are all in order to enlarge the rule-group.

(2).When the subsystems cooperate each other:

$$\frac{1}{k'}[(ab)+(a+b)] = J_2. \quad (12)$$

In this case the cooperation-coefficient k' is direct ratio with $J_2$. When $a=b=1/2$ and $k=1$, $J_2 =5/4$ is the maximum. Since the most increase is only 1/4, but the least decrease may reach to 3/4, it can show quantitatively that any object is destroyed easily and is constructed hard.

Further, humanity as an inseparable whole on Earth possesses common environment and benefit. A responsible idea for humanity should be set up, since every behavior of humanity correlates closely.

4.Theorem of transformation from quantity to quality on social development. It should be a direction of social development in the new century. We must discard single linear expansion, and a developing aim should transform into multi-levels, high quality and variety for human overall development.

We may research some concrete theories and problems in social system. Synergetics between man and nature is a large quantitative system theory on sustainable development [8,9], which is related with Chinese Yi-Jing and Lao-Zhuang philosophy. Here humanity and whole nature coexist equally and develop cooperatively. Scientific base of democracy expounds that democracy is a way of unblocked information and an open system, and is a process from random and chaos, through fluctuation and bifurcation, to order and consistency, and is a base that the modern society cooperates, then it will form the self-organized structures. Synergetics on war discuss the basic synergetic principle, the slaving principle and the order-parameter combining some war-examples. Moreover, mechanism of social synergy is a pluralistic structure, from which the social system may be stable. In a word, whole society must cooperation in order to exist and develop for humanity.

Based on the synergetic equations and its application, we proposed the equations on the rule of law. From these equations we proved mathematically that a society of the rule of law cannot lack any aspect for three types of the legislation, the administration and the judicature. Otherwise, we proposed an equation of corruption, and discussed quantitatively some threshold values for a system into corruption. From synergetics we obtain the Lorenz model, which may be a visualized two-party mechanism as a type of stable structure in democracy. A developed direction of society should be the combination from macroscopic to microscopic order, from an actual capable handling to an ideal pursuance [10].

**2. Social physics**

Usual social physics (sociophysics) is the application of the concepts of physics in the social sciences. Some basic conceptions of physics, for example, force and energy, etc., were applied to society. Social physics is the term first coined by Auguste Comte to describe the synoptic vision of the unity of all science, social and physical. He discussed the social statics and the social dynamics. The modern physics has introduced many new theories and new methods, for instance, phase transition, chaos, dissipative structure, fractals, etc. They not only have mathematical characters, but also possess very high universality.

Stewart proposed the suggested principles of social physics [11], and summarized the development of social physics [12]. T.M.Porter (1981) discussed a statistical survey of gases in Maxwell social physics. E.S.Knowles (1983) studied social physics, social gravity models and the effects of audience size and distance, etc. S.P.Restivo (1985) researched the social relations of physics, mysticism and mathematics. D.A.Walker researched economics as social physics in The Economic Journal (1991). Bernard and Killworth searched social physics for social network knowledge and theory [13]. Urry explored the increasing overlaps between sociology and physics, and explained the so-called small world phenomenon and corresponding new social physics [14]. Warntz discussed transportation, social physics and the law of refraction [15]. B.Hillier (2005) discussed the relations between social physics and phenomenology for cities as large physical objects animated and driven by human behaviour. Glymour searched social science and social physics [16]. G.D.Snooks (2007) discussed self-organization or selfcreation from social physics to realist dynamics, which can be employed to explain and predict the emergence of social structures, even of history itself.

Any system can be classified into two types: an open system exchanging matter or energy or information with its surroundings, and an isolated system doing not [17]. Of course, both are distinguished relatively for different regions and levels. It is consistent with Bergson viewpoint and with the theory of dissipative structure.

A complex social system is a set composed of many objects or elements with a certain relation, this will have a high-dimensional space of many variables. By using the method of the modern physics, it may be projected on a low-dimensional space of few variables. Further, various main quantities are classified into: 1.The random variable $\xi$ which cannot be dominated; 2.The relatively stable invariable in a certain region is called an extensive energy. 3.The extensive entropy which can describe various order degrees, organized powers, managed levels, and irreversibility, etc., and includes information. 4.The order parameter determining transitions of systems.

Based on above analysis we can discuss the eight aspects:

1.The social kinetic energy K and the social potential energy U.

2.The social force F. It is defined by interaction between the social elements

$$F = k\frac{dU}{dl}. \tag{13}$$

Here $l$ is an extension of the Bogardus-Simmel-Park social distance, or by

$$F = \frac{C}{\sqrt{K}}\frac{dK}{dt}. \tag{14}$$

Here C is the structure constant of a social system.

3.The social phase space. Its dimension equals independent of degree of freedom which describes essential character of a system. We consider that philosophy should be three dimensional for complex systems, so it may possess structure, stability, complexity and nonlinearity.

4.The social field. In the phase space a social field can be formed, whose function is

$$y = f(\xi_i, E_k, S_j, t). \tag{15}$$

For the average field

$$\bar{y} = \bar{f}(E_k, S_j, t). \tag{16}$$

The social field changes to pass through a critical threshold value, which may produce different structures of phase transition. This is a unification of necessity and fortuity.

5.The social quantum theory. In the society and history many quantities (country, man and so on) are quantized under distinguishable conditions of macroscopic parameter. It can be related with the extensive quantum theory [18-20].

6.The social theory of dissipative structures [17]. For a system, when it is open and the exchanged quantity reaches a threshold value, a new ordered structure can be formed. And if this quantity is provided sufficiently and constantly, such a stable self-organization can be maintained.

7.The social chaos. The nonlinear interactions among elements make usually system depended on initial conditions sensitively. When the order parameter reaches a certain vale, chaos appears, so economy collapses, war breaks out and so on.

8.The social fractals. It describes that the system possesses the self-similarity under some scaling transformations in a certain region. For example, some systems, the sizes of which are different, while whose structures are similar, and possess the same function. In this case the mathematics may apply the scale theory and the renormalization group.

**3.The social thermodynamics and research on social complex systems**

The social thermodynamics is a part of the social physics [21,22]. Ilya Prigogine proposed order through fluctuation for self-organization and social system. Lepkowski discussed the social thermodynamics of Ilya Prigogine [23]. Reed and Harvey discussed complexity and realism in the social sciences, and critical philosophy and non-equilibrium thermodynamics [24]. J.L.R.Proops (1987) researched entropy, information and confusion in the social sciences. K.D.Bailey (1990, 1993,1994) discussed social entropy theory and its application of nonequilibrium thermodynamics in human ecology, and living systems theory. G.A.Swanson, K.D.Bailey and J.G.Miller (1997) discussed social entropy and money in a living systems theory perspective. Balch researched hierarchic social entropy for an information theoretic measure of robot group diversity [25]. E.L.Khalil (1995) studied nonlinear thermodynamics and social science modeling. Scafetta, et al., studied concretely the thermodynamics of social processes for the teen birth phenomenon [26]. Zagreb researched an approach to a quantitative description of social systems based on thermodynamic formalism [27]. Stepanic, et al., described social systems using social free energy and social entropy [28]. Statistical physics as the fruitful framework to describe phenomena, recent Castellano, et al., studied collective phenomena emerging from the interactions of individuals as elementary units in social structures of social dynamics, and emphasized a comparison of model results with empirical data from social systems [29].

Since a state of single element in any complex social and natural systems is indeterminate and fluctuated, we discussed the social thermodynamics, which analyse mainly that the total system agrees with the statistical rules, and apply these methods which are analogous with thermodynamics and statistics.

In the social thermodynamics the social temperature may be defined by $T = c\overline{K}(t)$, where $\overline{K}$ is an average value of the social kinetic energy. While the social entropy may be defined by dS=dU/T. The present sociology has applied the statistical method, we research the universal statistical theory and principles in sociology. There are correspondingly social critical phenomena and phase transition. The threshold value at a critical point may pass through fluctuation to obtain a bifurcation point, which can produce different results of phase transition. It is a unification of necessity (a certain condition, threshold value) and fortuity (fluctuation, bifurcation point). Its mathematics may apply the catastrophe theory.

For any social system obeying statistical rule, we proposed the five fundamental laws and a set of relation:

The zeroth law: State fluctuation law. It originates from the dynamic background of system and the indeterminacy of state of single element, from this the fluctuation property in society and the social wave-motion are derived.

The first law: Extensive energy E conservation law. The extensive energy corresponds to the internal energy. For instance, all of population, natural resources, fund, land, and time, etc., are fixed. Earth is only one. These fundamental facts are neglected usually. In an isolated system dE/dt=0.

The second law: Extensive entropy S change law. S=klnW, where W is number of possible states of all element in this system. The extensive entropy is connected with the effective free energy. Usually it increase in an isolated system, and may decrease or increase in an open system.

The third law: Threshold transition law. If the extensive entropy reaches a maximum value in an isolated system, or exchanged quantity reaches a threshold value in an open system, the order parameter will change suddenly, and then system will exhibit a phase transformation, and will form a new state. It will be able to be an ordered dissipative structure, or a chaos state, or a more disordered, or an ordered but dead state which is similar with crystal. This is a bifurcation point which may produce different results. The fluctuations have an enlargement effect for reach of the threshold value.

The fourth law: Exclusion or inclusion law. The two different relations, exclusion or inclusion, exist among elements of various systems, both may be obvious or hidden forms. A synergetic relation can be formed only for inclusion elements. The causes of all conflicts and wars are exclusion of various benefits or powers.

The five laws are, respectively, random factor which cannot be controlled exactly, objective conditions which cannot be changed arbitrarily, basis of potentialities which may be exploited fully, key problems which must closely attach importance, and objects of study which should be distinguished strictly.

**4.Different relations among social elements and applications of the nonlinear sociology**

According to the different interrelations, which correspond to association, symbolic interactionism, conflict and so on in the modern sociology, among elements in the system, we may determinate a set of relations:

1.When elements have different characters and statistics, there is exclusion or inclusion each other among elements.

2.Among elements there can be the competitiveness (common restraint) or the cooperativeness (common promotion) which is the same with P.Kropotkin viewpoint. Both correspond to Park four social processes.

3.The relations among elements may be linear or nonlinear, the latter can derive bifurcation and chaos, etc. Any linear function y=cx+b changes (i.e., differential) to derive $dy/dt = c$, which only is an equal change. The nonlinear relations can derive various complex evolutional patterns.

4.The interrelations may produce complementarity, intersection, structure and so on.

They form the whole world and its rules. These laws and relations may be applied to discuss various social problems, for instance, development of a man self, protection of ecosystem, creativity-thinking, the Gulf war, especially, some important doctrines in economics and management. It may apply the determinant-stochastic relations in historical events.

Because human develops blindly, now a series of serious crises, for example, population explosion, environmental pollution, consumption of exhausted resources, unable regeneration ecology has dried up, has press on towards whole humanity. These like the terrific spirits. A nightmare on decay and elimination of some old nations has got entangled with our Earth Village.

Since humanity has been faced with these crises, a sustainable development is proposed. All society begins to attach importance to a corresponding theory. Although some brilliant theories are proposed, for example, the system dynamics as a human opinion and corresponding mathematical base, theoretical level and philosophical intensions should be researched and developed continuously.

We consider that humanity and our total natural circumstances are a huge common system. It includes above very much interacting elements.

Synergetics, as a quantitative cooperation theory of different parts of a system, may be considered as a strategy to copy with complex systems. Therefore, it can be applied to the sustainable development.

The basic equations of synergetics are the equations of single-mode laser [1,4]:

$$db/dt = -kb - i\sum g_\mu \alpha_\mu + F(t),$$
$$d\alpha_\mu/dt = -\gamma\alpha_\mu + ig_\mu * b\sigma_\mu + \Gamma_\mu(t), \quad (17)$$
$$d\sigma_\mu/dt = \gamma_{11}(d_0 - \sigma_\mu) + 2i(g_\mu\alpha_\mu b * -c.c.) + \Gamma_{\sigma,\mu}(t).$$

By the adiabatic approximation, these equations may be simply to the two equations,

$$db/dt = -kb + (g^2/\gamma)b\sum \sigma_\mu + F(t),$$
$$d\sigma_\mu/dt = \gamma_{11}(d_0 - \sigma_\mu) - (4g^2/\gamma)b*b\sigma_\mu. \quad (18)$$

When F(t)=0 and $\psi = b*b$, Eqs.(17) can become the Lotka-Volterra equations. The solution of the equations is cycle model with period. This corresponds to a circulation of natural resources. By using the slaving principle, the equations (18) can become the logistic equation:

$$d\psi/dt = aE\psi - \psi^2. \quad (19)$$

Its solution is:

$$\psi = \frac{aE}{1 + c\exp(-aEt)}. \quad (20)$$

which can be a $\Gamma$-form. This corresponds to increasing limit in economics of natural resources, etc. Moreover, based on the master equations (1), Eqs.(18) and (19) may also be derived [1,4].

Further, Eqs.(17) can extend to:

$$dx/dt = ax + bxy - cxz, \quad (21)$$

$$dy/dt = ey - fyz + gyx, \quad (22)$$

$$dz/dt = hz + kzx - lzy. \quad (23)$$

In various cases, dz/dt=0 corresponds to the symbiosis model in the population-dynamics, dx/dt=0 corresponds to the competition model, and dy/dt=0 corresponds to the cycle model and the predator-prey relation. The equations may derive various limit cycles.

In our human ecology, the Lotka-Velterra equations may be a zero-solution. For example, when water has dried up in a region, humanity cannot live. When the chaotic motion occurs, the formerly stable system is destabilized.

In the synergetics between humanity and nature [9], the order parameters are the threshold value of reaching limit and the critical point of circulationable state. We research a theory on the sustainable development, so the cooperativeness must be emphasized. The latter may also be applied to a modern mathematical representation on common restraint or promotion of the five elements in the Chinese traditional philosophy.

They show the two stages of development: 1. The moderate degree and limit on increase. 2. A good circulation on development. For a society of system dynamics, its essential characters should be orderly and regular, and corresponding, in the finally an ecological balance will be achieved.

They are also that the system dynamics must solve two necessarily fundamental problems.

The social thermodynamics is combined with the entropy, it will be obtained that the sustainable development should be connected with the moderate degree on the entropy production in systems and on the input negative entropy flow for open systems. Both correspond to the two solutions.

The three equations (21)~(23) correspond to the three elements on Sky-Earth-Man in Chinese traditional philosophy. Lao-tzu, a chinese ancient philosopher, said: Man models oneself on land, land models oneself on sky, sky models oneself on the Dao (law), Dao models oneself on nature. Lao-Zhuang philosophy thought: Man should be harmonious with environment and nature, both forms a suitable circle, finally it will achieve the highest goal of unifying humanity-nature. In this case the highest principle is not competition each other, and is a cooperation not only among various nations or countries, but among humanity and other plants and animals, humanity and total nature.

Our outlets are enterprising new energy sources, extending new living space and useful regions for humanity, expending new technology, and form a great industrial circulation like an agricultural ecological village. A chinese traditional poem described vividly the circulation:

Fallen flowers are not merciless and useless

They transform manures, and enrich flowers.

It is also an ideal state of Buddhism: without differences between life and death, and between me and world.

**5. The evolutional equation in system**

From above viewpoints, we propose the evolutional equation of system, whose one dimensional nonlinear evolutional equation of society is [8,4]:

$$dS/dt = \alpha S^m - \beta S^n + F(t). \tag{24}$$

Its different characters and solutions correspond to above various relations. If the evolutional equation is combined with the entropy, it will be obtained that the sustainable development should be connected with moderate degree on the entropy production in systems and on the input negative entropy flow for open systems [4,30].

For a linear case, m=n=1 and a stochastic factor F(t)=0, so Eq.(24) becomes

$$dS = \alpha S dt - \beta S dt. \tag{25}$$

Let S is entropy of system, $\alpha S dt$ includes the entropy produce $dS_i$ and positive entropy flow $dS_e^+$ for open systems, and $\beta S dt$ is an input negative entropy flow $dS_e^-$, so

$$dS = dS_i + dS_e^+ - dS_e^-. \tag{26}$$

Here $dS_i \geq 0$ is the entropy production inside the system, and $dS_e = dS_e^+ - dS_e^-$ is the entropy flow, which may be positive or negative. The total entropy can decrease when input entropy flow is negative. Therefore, the maximum entropy is

$$S_{max} = S_0 + dS_i + dS_e^+ \geq S_0 \geq dS_e^- > 0. \tag{27}$$

The maximum entropy defines a quantitative range of moderate degree on input negative entropy flow for any open system. Its absolute value is always greater than zero, but the total entropy can never become negative. This is namely a quantitative region of the moderate degree on an input negative entropy [30].

When $\beta=0$ and m=1, Eq.(24) is the Langevin equation. For the stochastic factor F(t)=0, we obtain the education equation [4]:

$$\frac{dE}{dt} = a(t)E . \qquad (28)$$

Its solution is:

$$E = E(0)\exp[\int a(t)dt] . \qquad (29)$$

This shows that the social benefit E is an exponential relation with the educative outlay $a$(t) devoted directly.

We proposed the confidence relations and the corresponding influence functions that represent various interacting strengths of different families, cliques and systems in organization by using the similar formulas of the preference relation and the utility function. In an economic system they can affect products, profit, prices and so on, such the system can produce a multiply connected topological economics [31]. When the changes of the product and the influence are independent one another, they may be a node or saddle point. When the influence function large enough achieves a certain threshold value, it will form a wormhole with loss of capital.

Moreover, Zhang Ming and I proposed and discussed money complex [32]. We researched possible decrease of entropy in some social systems [30]. The Chinese history shows a statistical rule, and forms a super-stable structure. In the Chinese traditional culture the rich and full moral educations are the thought basis of this social structure and manage state by morals. The Chinese select and supervise systems on various functionaries and the bureaucracy, for example, an imperial examinations system which accords with a theory of the dissipation structure, provide the govern base. The basis of the modern stable and harmonious society is equality. The comparison and combination at all times and in all over the world will redound to build a perfect society [33].

We try to put forward new understand on civilization relative to barbarism: The civilization is a rational and nonviolent mode to solve much widespread difference in human society [34]. This connotation of civilization is extensively existence in economical, political, cultural activities and in daily life. This has some similar characters with universal civilization. But, it should be different with usual civilization, which possessed a characteristic of the diversity. We discussed the value, perplexity, current and so on of this connotation of civilization. Such we demurred to a known book 《The Clash of Civilizations and the Remaking of World Order》[35].

From this brief introduction we can perform some quantitative calculations. It overcomes the infantilism of the social physics in 18[th] century, and gives a conception of the modern physics on mankind society in a thought space, and set up a bridge between physics and social systems.

# bibliography

4.Yi-Fang Chang, Social Synergetics. Beijing: Science Press. 2000.
5.W.Weidlich, Phys.Reports. 204,1(1991).
6.Yu Xie, Sociological Methodology and Quantitative Research. Social Science Academic Press (China). 2006.
7.J.S.Bell, Physics. 1,195(1964).
8.Yi-Fang Chang, Equation of social evolvement and synergetic theory on reformatory of Chinese economy. Theory and Application of System Science. Sichuan University Press. 1996.234-237.
9.Yi-Fang Chang, Knowledge economy and synergetics between man and nature. Selection of Knowledge Economy in China. Chinese Economy Press. 2000.275-278.
10.Yi-Fang Chang, Synergetic Application, Equations on Rule of Law and Two-Party Mechanism. arXiv:0809.0335.
11.J.Q.Stewart, Science. 106(2748),179(1947).
12.J.Q.Stewart, Amer.J.Phys. 18,239(1950).
13.H.R.Bernard and P.Killworth, Connections. 20,16(1997).
14.J.Urry, Global Networks. 4,109(2004).
15.W.Warntz, The Professional Geographer. 9(4),2(2005).
16.C.Glymour, Behavioral Science. 28(2),126(2007).
17.Yi-Fang Chang and Hao Che, Theory of dissipative structures and scientific sociology. Science & Technology Herald. 3,54(1987).
18.Yi-Fang Chang, J.Yunnan University. 15(4),297(1993).
19.Yi-Fang Chang, Physics Essays. 15(2),133(2002).
20.Yi-Fang Chang, J.Rel.Psych.Res. 27(4),190(2004).
21.S.Galam,Y.Gefen and Y.Shapir, J.Math.Sociol. 9,1(1982).
22.P.Ball, Physica. A314,1(2002).
23.W.Lepkowski, Chemical and Engineering News. 57,30(1979).
24.M.Reed and D.L.Harvey, Journal for the Theory of Social Behaviour. 22,353(1992).
25.T.Balch, Autonomous Robots. 8(3),209(2000).
26.N.Scafetta, P.Hamilton and P.Grigolini, The Thermodynamics of Social Processes: The Teen Birth Phenomenon. arXiv:cond-mat/0009020(2000).
27.C.Zagreb, Entropy, 2,98(2000).
28.J.Stepanic, G.Sabol and M.S.Zebec, Kybernetes. 34(6),857(2005)
29.C.Castellano, S.Fortunato and V.Loreto, Rev.Mod.Phys. 81,591(2009).
30.Yi-Fang Chang, Moderate Degree of Input Negative Entropy Flow and Decrease of Entropy in Astrophysics, Biology, Psychology and Social Systems. arXiv:0910.0649.
31.Yi-Fang Chang, Multiply Connected Topological Economics and Confidence Relation arXiv:0711. 0235.
32.Yi-Fang Chang, Ming Zhang, Science (China). 5,57(1998).
33.Yi-Fang Chang, Research on super-stable structure and its thought culture of Chinese traditional society. Journal of Hunan City University. 30,3,13-19(2009).
34.Ping Zhang and Yi-Fang Chang, Consideration and Reunderstand on Civilization. Social Science Research. 2,68(2008).
35.S.P.Huntington, The Clash of Civilizations and the Remaking of World Order. Georges Borchardt, Inc. 1996.
11